\documentclass[12pt,english,aps,secnumarabic,nofootinbib,groupedaddress,notitlepage]{revtex4}

\usepackage[T1]{fontenc}
\usepackage[utf8]{inputenc}
\usepackage{babel}
\usepackage{amsmath}
\usepackage{amssymb}
\usepackage{esint}
\usepackage[unicode=true,pdfusetitle,
 bookmarks=true,bookmarksnumbered=false,bookmarksopen=false,
 breaklinks=false,pdfborder={0 0 1},backref=section,colorlinks=false]
 {hyperref}
\hypersetup{
 citecolor=blue}

\makeatletter
\@ifundefined{textcolor}{}
{%
 \definecolor{BLACK}{gray}{0}
 \definecolor{WHITE}{gray}{1}
 \definecolor{RED}{rgb}{1,0,0}
 \definecolor{GREEN}{rgb}{0,1,0}
 \definecolor{BLUE}{rgb}{0,0,1}
 \definecolor{CYAN}{cmyk}{1,0,0,0}
 \definecolor{MAGENTA}{cmyk}{0,1,0,0}
 \definecolor{YELLOW}{cmyk}{0,0,1,0}
}

\usepackage{babel}

\usepackage{amsfonts}\usepackage{bbm}\usepackage{euscript}
\usepackage{dcolumn}
\usepackage{bm}
\usepackage{epsfig}\epsfclipon
\usepackage{slashed}

\DeclareMathOperator{\Unitary}{U}
\DeclareMathOperator{\SO}{SO}
\DeclareMathOperator{\Tr}{Tr}

\makeatother

\begin{document}

\title{Basis Tensor Gauge Theory}

\author{Daniel J. H. Chung}

\email{danielchung@wisc.edu}

\affiliation{Department of Physics, University of Wisconsin-Madison, Madison,
WI 53706, USA}

\author{Ran Lu}

\email{rlu@wisc.edu}

\affiliation{Department of Physics, University of Wisconsin-Madison, Madison,
WI 53706, USA}
\begin{abstract}
We reformulate gauge theories in analogy with the vierbein formalism
of general relativity. More specifically, we reformulate gauge theories
such that their gauge dynamical degrees of freedom are local fields
that transform linearly under the dual representation of the charged
matter field. These local fields, which naively have the interpretation
of non-local operators similar to Wilson lines, satisfy constraint
equations. A set of basis tensor fields are used to solve these constraint
equations, and their field theory is constructed. A new local symmetry
in terms of the basis tensor fields is used to make this field theory
local and maintain a Hamiltonian that is bounded from below. The field
theory of the basis tensor fields is what we call the basis tensor
gauge theory.
\end{abstract}
\maketitle

\section{Introduction}

Gauge theories (see e.g. \cite{Weyl:1919fi,Weyl:1929fm,Yang:1954ek,Abers:1973qs,Itzykson:1980rh,Polyakov:1987ez,'tHooft:1995gh,Weinberg:1996kr})
are extremely robust and successful in describing fundamental interactions
of nature such as in the Standard Model (SM) of particle physics \cite{Glashow:1961tr,Weinberg:1967tq,Salam:1968rm,Gross:1973id,Politzer:1973fx,Weinberg:1996kr,Ramond:1999vh,Langacker:2010zza,Aad:2012tfa,Chatrchyan:2012xdj}.
In the usual gauge theoretic formulation, the gauge field is a connection
on principal bundles (see e.g.~\cite{Nakahara:2003nw,Wu:1975es}).
In the usual formulation of general relativity, Christoffel symbols
are connections on the tangent bundle and can be expressed nonlinearly
in terms of the metric. Another widely used formulation of general
relativity that is particularly useful when spinors need to be defined
in curved spacetime is the vierbein formalism. In this formalism,
$N$ basis vector fields are introduced as a way of taking the square
root of the metric, in which $N$ is the dimension of spacetime. However,
in the case of gauge theories, there is no widely known analogous
vierbein formulation, presumably because there is no obvious nontrivial
metric analog that carries the gauge field information. In this work,
we construct a vierbein-like field theory of a $\Unitary(1)$ gauge
theory coupled to complex scalars.

Our approach is to construct an explicit representation of the matter
field direction in the group representation space as a \emph{local}
Lorentz tensor field that is subject to constraints that arise from
matching to the ordinary gauge field connection. This tensor transforms
as a dual to the matter representation and the constraint equation
is reminiscent of the relationship between the spacetime vierbein
field and the Christoffel symbol. In this sense, this tensor field
is the analog of the general relativistic vierbein for our construction.
We then solve this constraint equation by decomposing the log of this
tensor in terms of $N$ fields that we call basis tensor fields. These
fields effectively span the Lie algebra that generates the tensor
field. The field theory of these basis tensor fields is local and
has new local symmetries that allow this theory to perturbatively
match to ordinary gauge theories.

More explicitly, the vierbein-like field is taken to be a Lorentz
tensor $G(x)$ that satisfies a constraint equation. Since $G(x)$
transforms in the gauge group representation space as a dual to the
matter field, if a matter field $\phi$ is charged under $\Unitary(1)$
with charge 1, $G(x)$ transforms with a charge -1 and the object
$\phi(x)G(x)$ is gauge invariant. We show that the minimal Lorentz
tensor rank of $G(x)$ that has this desired dual property and can
accommodate the local gauge field degrees of freedom is 2: i.e. $G_{\,\,\,\,\beta}^{\alpha}(x)$.
The constraint equation of $G_{\,\,\,\,\beta}^{\alpha}(x)$ can be
solved in terms of another set of unconstrained fields $\{\theta^{a}(x)(H^{a})_{\,\,\,\,\,\nu}^{\mu}|a\in\{0,...N-1\}\}$
(similarly in spirit to sigma model constructions), which are the
basis tensor fields. The field theory of $\theta^{a}(x)$ is what
we will call \emph{basis tensor gauge theory} (BTGT) and is an alternate
to the gauge theory description in terms of $A_{\mu}(x)$. It is the
theory of $\theta^{a}$ that will exhibit a \emph{new local symmetry}
to maintain the (perturbative) isomorphism between the usual gauge
theory and BTGT. 

Giving a vierbein expression of gauge fields in this work makes gauge
theories look more like general relativity, which in some sense is
similar in philosophy to Kaluza-Klein theories \cite{Overduin:1998pn},
but the approach here is different in that we try to minimize the
disturbance to the theory. More precisely, instead of trying to unify
the gauge theory with spacetime dynamics, the theory is merely rewritten
such that the gauge fields more closely resemble the matter fields.
In the usual model building description of gauge theories, the gauge
fields are put on a different footing than the matter fields in that
the gauge fields do not form a linear representation of the gauge
group while the matter fields typically do. In our approach, the $G_{\,\,\,\,\beta}^{\alpha}(x)$
fields, which have the same information as the gauge fields, form
a linear representation. The most interesting result arising from
this is the emergence of a local symmetry that is independent of the
ordinary gauge symmetry.

To our knowledge, the previous work that most closely resembles our
approach is that of Mandelstam \cite{Mandelstam:1962mi}, in which
the group space linear representation is given as an object similar
to a Wilson line (for several examples of the vast literature on this
topic, see e.g.~\cite{Wilson:1974sk,Giles:1981ej,Migdal:1984gj,Terning:1991yt,Gross:2000ba,Kapustin:2005py,Cherednikov:2008ua}
and references therein). In some sense, this object can be viewed
as the analog of $G_{\,\,\,\,\beta}^{\alpha}$.%
\footnote{A Wilson line transforms as a non-local adjoint. If one views one
end of the Wilson line to be at infinity and demands that the gauge
transformations vanish there, then it looks as if the Wilson line
transforms as a fundamental.%
} However, in addition to the fact that Wilson lines are manifestly
non-local, the purpose of Mandelstam's work was to formulate gauge
theories without any gauge fields. In contrast, the purpose of our
work is to explicitly construct a gauge group representation direction
as a local spacetime Lorentz tensor field, not to hide the group representation
space.

The order of presentation is as follows. In Sec.~\ref{sec:A-group-space},
we derive the relationship between $G_{\,\,\,\,\,\beta}^{\alpha}(x)$
and $A_{\mu}(x)$ using an ansatz analogous to the equivalence principle.
This relationship serves as a constraint equation. We then solve this
constraint equation using the basis tensors $\theta^{a}H^{a}$. In
Sec.~\ref{sec:thetaisintegralofA}, we show how the integral over
$A_{\mu}$ is related to the $\theta^{a}$ field. The naive non-locality
will be eliminated by the symmetries imposed when defining the partition
function in Sec.~\ref{sec:Path-Integral}. In Sec.~\ref{sec:Elementary-computation},
we go through the exercise of constructing a BTGT model based on the
recipe in Sec.~\ref{sec:Path-Integral}. We give Feynman rules and
apply them to a a simple scattering computation. Sec.~\ref{sec:Peculiarities-of-the}
lists some of the peculiarities of the model: a) each charged elementary
field has its own group direction field (that are all related to each
other through the same $\theta^{a}$) and the covariant derivative
can be written as a peculiar divergence of a composite field; b) the
Hamiltonian is bounded from below despite the fact that the $\theta^{a}$
theory is a higher derivative theory; c) BTGT gives a novel way of
computing non-local correlators. We conclude by speculating on future
research directions. The Appendices present explanations of the minimal
rank of the Lorentz tensor for BTGT as well as the relationship of
the new local symmetry to translational symmetry. The last appendix
section explicitly displays the analogy between $G_{\,\,\,\,\,\beta}^{\alpha}(x)$
and the general relativistic vierbein.

\section{\label{sec:A-group-space}A group space matter direction field }

The purpose of this work is to construct an alternate description
to the usual gauge field that puts matter fields and the gauge fields
on a more similar mathematical categorization. Because relativistic
quantum field theory naturally partitions into relativistic tensor
field degrees of freedom, any alternate local description of the gauge
field has a natural description in terms of Lorentz tensors. We therefore
define a \emph{local Lorentz tensor} field in the dual representation
of the matter field which describes the ``direction'' of the matter
field in the group representation space. For simplicity, we focus
here on the $\Unitary(1)$ group, although we foresee no insurmountable
obstacles to generalize this to non-Abelian theories.

Given a field $\phi$ that is a complex scalar charged under $\Unitary(1)$
as 
\begin{equation}
\phi(x)\rightarrow e^{i\theta(x)}\phi(x),
\end{equation}
we wish to construct a Lorentz tensor object $G_{\alpha\beta}$ and
its field theory that exhibits the $\Unitary(1)$ gauge group transformation
property 
\begin{equation}
G_{\,\,\,\,\,\beta}^{\alpha}(x)\rightarrow G_{\,\,\,\,\,\,\beta}^{\alpha}(x)e^{-i\theta(x)},\label{eq:Gtransform}
\end{equation}
such that $\phi G_{\,\,\,\,\,\beta}^{\alpha}$ is gauge invariant.
We note that we can view $G_{\,\,\,\,\,\beta}^{\alpha}$ as the direction
in gauge group linear representation space. We discuss in Appendix
\ref{sec:Lower-rank-tensor} that a rank 2 Lorentz tensor is the smallest
rank for which such a local description alternate to the gauge field
is possible. We also show in Appendix \ref{sec:vierbein-analogy}
how $G_{\,\,\,\,\,\beta}^{\alpha}(x)$ is analogous to the general
relativistic vierbein. To construct the theory of $G_{\,\,\,\,\,\,\,\beta}^{\alpha}$,
we will match to the known $A_{\mu}$ gauge theory. To this end, we
need to find a relationship between $G_{\,\,\,\,\,\beta}^{\alpha}$
and the ordinary gauge field $A_{\mu}$.

Some degree of rigidity in the construction can be attained and the
spirit of making gauge theories look more like general relativity
can be followed if we use an analog of the equivalence principle approach
(see e.g. \cite{Weinberg:1972kfs}) of making a general coordinate
transformation away from the freely falling frame of the matter to
define the Christoffel symbol (the connection on the tangent bundle).%
\footnote{Of course, this is simply an ansatz for defining the representation
since there is no universality of charge to mass ratio in gauge theories.%
} Here the analog of the freely falling frame can be defined to be
the frame in which the $\Unitary(1)$ connection $A_{\mu}(x)$ vanishes
at a spacetime point $x_{1}$, since $A_{\mu}$ enters without any
derivatives in the matter Lagrangian: 
\begin{equation}
\mathcal{L}_{\phi}=(\partial_{\mu}+iA_{\mu})\phi(\partial^{\mu}-iA^{\mu})\phi^{*}.\label{eq:cov}
\end{equation}
(Note that this definition is in contrast with the gravitational equivalence
principle which relies on the equation of motion rather than the Lagrangian.)
In this frame, the Lagrangian at point $x_{1}$ looks like there is
no gauge field (just as locally, the Christoffel symbol vanishes in
the freely falling frame): 
\begin{equation}
\mathcal{L}_{\phi}(x_{1})=\partial_{\mu}\tilde{\phi}\partial^{\mu}\tilde{\phi}^{*}(x_{1}).
\end{equation}
We demand in this special gauge frame that the vierbein-like tensor
field has the following value at point $x_{1}$: 
\begin{equation}
\tilde{G}_{\alpha\beta}(x_{1})=S_{\alpha\beta}(x_{1}).\label{eq:specialframe}
\end{equation}
Upon making a gauge transformation to move to the general frame, we
have 
\begin{equation}
\phi(x)=\tilde{\phi}(x)e^{i\theta(x)},\label{eq:phaseshift}
\end{equation}
which gives 
\begin{equation}
\mathcal{L}_{\phi}(x_{1})=(\partial_{\mu}-i\partial_{\mu}\theta)\phi(\partial^{\mu}+i\partial^{\mu}\theta)\phi^{*}.
\end{equation}
Comparing this to the usual definition of the connection covariant
derivative 
\begin{equation}
D_{\mu}=\partial_{\mu}+iA_{\mu}
\end{equation}
that appears in Eq.~(\ref{eq:cov}), we identify the connection as
\begin{equation}
A_{\mu}(x_{1})=-\partial_{\mu}\theta(x_{1}).\label{eq:defofgaugepotential}
\end{equation}
Note that this does not mean $A_{\mu}(x)$ is a pure gauge configuration
everywhere, as this equation applies at only one point $x_{1}$. Since
$G_{\alpha\beta}$ is \textbf{defined} to obey the transformation
rule of Eq.~(\ref{eq:Gtransform}), we have 
\begin{equation}
G_{\alpha\beta}(x_{1})=S_{\alpha\beta}(x_{1})e^{-i\theta(x_{1})}.\label{eq:rank2}
\end{equation}
Because of Eq.~(\ref{eq:defofgaugepotential}), we want to solve
for $\partial_{\mu}\theta(x_{1})$ in terms of $G$ evaluated at $x_{1}$.
To achieve this, we take derivatives of the general gauge-transformed
object 
\begin{equation}
e^{i\theta}(G^{-1})^{\alpha\beta}\partial_{\alpha}(G_{\beta\mu}e^{-i\theta})=(G^{-1})^{\alpha\beta}\partial_{\alpha}G_{\beta\mu}(x)-i\partial_{\mu}\theta(x)
\end{equation}
and evaluate this general expression at $x_{1}$ in the special gauge
frame: 
\begin{equation}
e^{i\theta}(\tilde{G}^{-1})^{\alpha\beta}\partial_{\alpha}(\tilde{G}_{\beta\mu}e^{-i\theta})|_{x_{1}}=(\tilde{G}^{-1})^{\alpha\beta}\partial_{\alpha}\tilde{G}_{\beta\mu}(x_{1})-i\partial_{\mu}\theta(x_{1}).
\end{equation}
Because of Eq.~(\ref{eq:defofgaugepotential}), we conclude 
\begin{equation}
A_{\mu}(x_{1})=-i\left[(G^{-1})^{\alpha\beta}(\partial_{\alpha}G_{\beta\mu})|_{x_{1}}-(\tilde{G}^{-1})^{\alpha\beta}(\partial_{\alpha}\tilde{G}_{\beta\mu})|_{x_{1}}\right],\label{eq:vecpotatpoint}
\end{equation}
in which 
\begin{equation}
G_{\alpha\beta}(x)\equiv\tilde{G}_{\alpha\beta}(x)e^{-i\theta(x)}
\end{equation}
is the general gauge field.

We can now simplify Eq.~(\ref{eq:vecpotatpoint}) further by noting
that Eq.~(\ref{eq:vecpotatpoint}) has an additional set of $\tilde{\Unitary}(1)$
symmetry transformations: 
\begin{equation}
G_{\alpha\beta}\rightarrow G_{\alpha\beta}e^{-i\Lambda_{\beta}(x)}\,\,\,\,\,\,\,\,\,\,\,\,\,\,\,\,\,\,\,\,\tilde{G}_{\alpha\beta}\rightarrow\tilde{G}_{\alpha\beta}e^{-i\Lambda_{\beta}(x)}
\end{equation}
that leaves Eq.~(\ref{eq:vecpotatpoint}) invariant. This means we
can use it to choose $\partial_{\alpha}\tilde{G}_{\beta\mu}=0$ as
follows. First, we execute a $\tilde{\Unitary}(1)$ transform to go
to the barred frame: 
\begin{equation}
(G^{-1})^{\alpha\beta}(\partial_{\alpha}G_{\beta\mu})=(\bar{G}^{-1})^{\alpha\beta}(\partial_{\alpha}\bar{G}_{\beta\mu})-i\partial_{\mu}\Lambda_{\mu}\,\,\,\,\,\,\,\,\mbox{no sum over }\mu
\end{equation}
\begin{equation}
(\tilde{G}^{-1})^{\alpha\beta}(\partial_{\alpha}\tilde{G}_{\beta\mu})=(\bar{\tilde{G}}^{-1})^{\alpha\beta}(\partial_{\alpha}\bar{\tilde{G}}_{\beta\mu})-i\partial_{\mu}\Lambda_{\mu}\,\,\,\,\,\,\,\,\mbox{no sum over }\mu
\end{equation}
where the yet-to-be-determined $\Lambda_{\mu}(x)$ parametrizes the
transformation to the barred frame. We can then impose the condition
\begin{equation}
(\bar{\tilde{G}}^{-1})^{\alpha\beta}(\partial_{\alpha}\bar{\tilde{G}}_{\beta\mu})=0\label{eq:gaugecondition}
\end{equation}
to solve for $\Lambda_{\mu}$. This implies that 
\begin{equation}
(\bar{G}^{-1})^{\alpha\beta}(\partial_{\alpha}\bar{G}_{\beta\mu})=(G^{-1})^{\alpha\beta}(\partial_{\alpha}G_{\beta\mu})-(\tilde{G}^{-1})^{\alpha\beta}(\partial_{\alpha}\tilde{G}_{\beta\mu}).
\end{equation}
In this $\tilde{\Unitary}(1)$ fixed system, we have 
\begin{equation}
A_{\mu}(x)=-i(\bar{G}^{-1})^{\alpha\beta}(\partial_{\alpha}\bar{G}_{\beta\mu}),
\end{equation}
in which the bar indicates that we have fixed the $\tilde{\Unitary}(1)$
gauge through Eq.~(\ref{eq:gaugecondition}). For notational convenience,
we can simply drop the bar: \emph{i.e.}, we then have 
\begin{equation}
\boxed{A_{\mu}=-i(G^{-1})_{\,\,\,\,\beta}^{\alpha}\partial_{\alpha}G_{\,\,\,\,\mu}^{\beta}}.\label{eq:mapaandG}
\end{equation}
This equation gives the relationship between the vierbein-like field
$G_{\,\,\,\,\,\beta}^{\alpha}$ and the gauge field $A_{\mu}$. We
note that because of the way the Lorentz tensor indices are contracted,
this is not a pure gauge configuration. As explained in Appendix \ref{sec:Lower-rank-tensor},
this is in contrast with the situation with lower rank tensors. It
is also here that we see how Eq.~(\ref{eq:mapaandG}) is reminiscent
of the relationship between the Christoffel symbol and the vierbein.
In the next section, we will introduce new basis fields to decompose
$G_{\,\,\,\,\,\beta}^{\alpha}$.

\section{Decomposing the vierbein}

In this section, we will show that demanding (a) the reality condition
implied by Eq.~(\ref{eq:mapaandG}), (b) that $G_{\,\,\,\,\beta}^{\alpha}$
transform like a $(1\,\,\,\,1)$ Lorentz tensor, and (c) $G_{\,\,\,\,\,\beta}^{\alpha}\rightarrow\eta_{\,\,\,\,\,\beta}^{\alpha}$
in the vacuum limit, ``uniquely'' fixes 
\begin{equation}
\boxed{G_{\,\,\,\,\,\beta}^{\alpha}\in\bigoplus_{n=1}^{N}\Unitary(1)}\label{eq:reduciblerep}
\end{equation}
where $N=4$ for four spacetime dimensions and each $\Unitary(1)$
in the sum means a 1 dimensional representation. Each of $N$ phase
fields are what we will call $\theta^{a}H^{a}$, which are the basis
tensors.

Since the $G_{\,\,\,\,\,\beta}^{\alpha}$ constrained by Eq.~(\ref{eq:mapaandG})
are difficult to work with, we will solve this constraint equation
here in terms of the unconstrained fields. Consider the representation
(just as in sigma model constructions) 
\begin{equation}
G_{\,\,\,\,\,\,\mu}^{\beta}=\left(e^{i\theta^{a}(x)H^{a}}\right)_{\,\,\,\,\,\,\mu}^{\beta}\label{eq:matrix}
\end{equation}
\begin{equation}
(G^{-1})_{\,\,\,\,\,\,\beta}^{\alpha}=\left(e^{-i\theta^{a}(x)H^{a}}\right)_{\,\,\,\,\,\,\beta}^{\alpha},
\end{equation}
in which $\theta^{a}$ is real without loss of generality, $H^{a}$
is a general set of constant matrices (maximally $2N^{2}$ such matrices
exist where $N=4$ for 4 spacetime dimensions), and the repeated indices
here are summed. We note that Eq.~(\ref{eq:matrix}) contains an
assumption about going to a manifestly Lorentz-invariant vacuum in
the limit of $\theta(x)\rightarrow0$; \emph{i.e.}, in the limit $\theta(x)\rightarrow0$,
$G_{\,\,\,\,\,\beta}^{\alpha}$ becomes an identity matrix, which
is Lorentz invariant. To satisfy Eq.~(\ref{eq:mapaandG}), we expand
for small $\theta$: 
\begin{equation}
A_{\mu}=\partial_{\alpha}\theta^{a}(H^{a})_{\,\,\,\,\,\mu}^{\alpha}+O(\theta^{2}).\label{eq:establishreality}
\end{equation}
This says that $H^{a}$ should be a \emph{real} matrix.

If we keep the entire power series, we have 
\begin{equation}
G_{\,\,\,\,\,\,\mu}^{\beta}=\left(\sum_{n=0}^{\infty}\frac{1}{n!}\left[i\theta^{a}(x)H^{a}\right]^{n}\right)_{\,\,\,\,\,\,\mu}^{\beta}.
\end{equation}
We can take any generic $m$ power term in this series as follows:
\begin{equation}
\theta^{a_{1}}\theta^{a_{2}}...\theta^{a_{m}}H^{a_{1}}H^{a_{2}}...H^{a_{m}}.
\end{equation}
We define each $H^{a}$ to transform like a rank 2 tensor under Lorentz
transformations. Hence, each such term transforms as 
\begin{equation}
\theta^{a_{1}}\theta^{a_{2}}...\theta^{a_{m}}\Lambda H^{a_{1}}\Lambda^{-1}\Lambda H^{a_{2}}\Lambda^{-1}...\Lambda H^{a_{m}}\Lambda^{-1},
\end{equation}
which means that the matrix ansatz Eq.~(\ref{eq:matrix}) does transform
like a $(1\,\,1)$ tensor under Lorentz transformations.

Let us now consider the reality condition on the rest of the terms
in the power series. First, we use the Baker-Campbell-Hausdorff formula
to express the gauge field in terms of a parametric integral: 
\begin{equation}
\partial_{\alpha}G_{\,\,\,\,\,\,\mu}^{\beta}=i\partial_{\alpha}\theta^{f}\int_{0}^{1}dt\left[e^{i(1-t)\theta^{a}H^{a}}H^{f}e^{it\theta^{a}H^{a}}\right]_{\,\,\,\,\,\,\mu}^{\beta}
\end{equation}
\begin{equation}
A_{\mu}=\partial_{\alpha}\theta^{f}\int_{0}^{1}dt\left[e^{-it\theta^{a}H^{a}}H^{f}e^{it\theta^{a}H^{a}}\right]_{\,\,\,\,\,\,\mu}^{\alpha}.
\end{equation}
Taking the complex conjugate of this yields 
\begin{equation}
\left(\partial_{\alpha}\theta^{f}\int_{0}^{1}dt\left[e^{-it\theta^{a}H^{a}}H^{f}e^{it\theta^{a}H^{a}}\right]_{\,\,\,\,\,\,\mu}^{\alpha}\right)^{*}=\partial_{\alpha}\theta^{f}\int_{0}^{1}dt\left[e^{it\theta^{a}H^{a}}H^{f}e^{-it\theta^{a}H^{a}}\right]_{\,\,\,\,\,\,\mu}^{\alpha}.
\end{equation}
Next, using the identity 
\begin{equation}
e^{A}Be^{-A}=\sum_{n=0}^{\infty}\frac{1}{n!}[A,[A,[...[A,B]]...]],
\end{equation}
we split even and odd powers 
\begin{align}
\partial_{\alpha}\theta^{f}\int_{0}^{1}dt\left[e^{it\theta^{a}H^{a}}H^{f}e^{-it\theta^{a}H^{a}}\right]_{\phantom{\alpha}{\mu}}^{\alpha} & =\partial_{\alpha}\theta^{f}\int_{0}^{1}dt\sum_{n={\rm odd}}^{\infty}\frac{1}{n!}\left[it\theta^{a_{n}}H^{a_{n}},\left[...\left[it\theta^{a_{1}}H^{a_{1}},H^{f}\right]...\right]\right]_{\phantom{\alpha}\mu}^{\alpha}\nonumber \\
 & +\partial_{\alpha}\theta^{f}\int_{0}^{1}dt\sum_{n={\rm even}}^{\infty}\frac{1}{n!}\left[it\theta^{a_{n}}H^{a_{n}},\left[...\left[it\theta^{a_{1}}H^{a_{1}},H^{f}\right]...\right]\right]_{\phantom{\alpha}\mu}^{\alpha}
\end{align}
to separate the sign dependence. Although the even power terms do
not depend on the sign in front of $it\theta^{a}H^{a}$, the odd power
terms are odd under the sign change. Since $\theta^{a}(x)$ and $\partial_{\alpha}\theta^{f}(x)$
can have any value, we conclude that the only representation for which
$A_{\mu}$ can be represented this way is 
\begin{equation}
\partial_{\alpha}\theta^{f}\left[\theta^{a_{2m+1}}H^{a_{2m+1}},\left[...\left[\theta^{a_{1}}H^{a_{1}},H^{f}\right]\right]...\right]_{\phantom{\alpha}\mu}^{\alpha}=0\label{eq:eachcommutatorvanishes}
\end{equation}
for every integer $m\geq0$. Hence, we conclude that the only matrices
$H^{f}$ that can satisfy this are $(1\,\,\,\,1)$ Lorentz tensors
that satisfy 
\begin{equation}
\boxed{\left[H^{a},H^{b}\right]=0}\label{eq:commutingmatrices}
\end{equation}
These form a reducible representation of $\Unitary(1)$ given by Eq.~(\ref{eq:reduciblerep}).

One explicit representation of $H^{a}$ is furnished by the following
real polarization vectors: 
\begin{equation}
\left(H^{a}\right)_{\,\,\,\,\,\nu}^{\mu}=\psi_{(a)}^{\mu}\psi_{(a)\nu},\label{eq:psibasis}
\end{equation}
in which 
\begin{equation}
\psi_{(a)}^{\mu}=\Lambda_{\,\,\,\,\,\, a}^{\mu}
\end{equation}
are components of the Lorentz transformation matrix $\Lambda$ (the
fundamental representation of $\SO(N-1,1)$). The $N$ fields 
\begin{equation}
\theta^{a}\left(x\right)\left(H^{a}\right)_{\,\,\,\,\,\nu}^{\mu}\,\,\,\,\,\,\,\,\mbox{no sum on }a
\end{equation}
appearing in Eq.~(\ref{eq:matrix}) span the spacetime tensor space
and can be used to expand the vierbein-like field $G_{\,\,\,\,\,\,\beta}^{\alpha}(x)$.
On the other hand, they span the Lie algebra of the gauge group instead
of the group representation itself. This makes them more like gauge
fields. The fact that $H^{a}$ is a complete basis is manifest in
the identities 
\begin{equation}
\sum_{a}H^{a}=\mathbb{I}
\end{equation}
\begin{equation}
\Tr\left(H^{a}H^{b}\right)=\delta^{ab}.
\end{equation}

We can summarize this section with the statement that the vierbein-like
field which transforms as a dual of the $\Unitary(1)$ matter representation
is given by Eq.~(\ref{eq:matrix}), in which the $H^{a}$ are real,
commuting $N\times N$ matrices that transform like a $(1\,\,\,\,1)$
Lorentz tensor.

\section{\label{sec:thetaisintegralofA}$\theta^{a}$ as an integral over
$A_{\mu}$ }

To gain intuition regarding the variable $\theta^{a}$, it is instructive
to express $\theta^{a}$ in terms of $A_{\mu}$. Since the $H^{a}$
are commuting matrices, Eq.~(\ref{eq:mapaandG}) gives 
\begin{equation}
\boxed{A_{\mu}=\sum_{a}\partial_{\alpha}\theta^{a}(H^{a})_{\,\,\,\,\,\mu}^{\alpha}}.\label{eq:establishreality-1}
\end{equation}
This equation can be solved for $\theta^{a}$: 
\begin{equation}
\boxed{\theta^{a}(y)=\int_{Y_{0}}^{y}dz^{\mu}(H^{a})_{\,\,\,\,\,\,\mu}^{\lambda}A_{\lambda}(x(z,y))+Z^{a}(y)}\label{eq:thetaa}
\end{equation}
\begin{equation}
x^{\lambda}(z,y)\equiv(H^{a})_{\,\,\,\,\,\,\mu}^{\lambda}z^{\mu}+\sum_{b\neq a}(H^{b})_{\,\,\,\,\,\,\,\,\nu}^{\lambda}y^{\nu},
\end{equation}
in which the $dz^{\mu}$ integral is over a straight path connecting
$Y_{0}$ and $y$, and the $Z^{a}(x)$ are the zero modes of the derivative
operator in Eq.~(\ref{eq:establishreality-1}) and satisfy 
\begin{equation}
(H^{a})_{\,\,\,\,\mu}^{\alpha}\frac{\partial}{\partial x^{\alpha}}Z^{a}(x)=0\,\,\,\,\,\,\,\mbox{no sum over }a.\label{eq:lowerdimgaugesym-1}
\end{equation}
This means $Z^{a}$ is a function that depends on a 3-dimensional
subspace of the 4-dimensional space. Another way of saying this is
that $Z^{a}(x)$ is translationally invariant: 
\begin{equation}
Z^{a}(y+T_{a}\psi_{(a)})=Z^{a}(y)\label{eq:zeromodedef}
\end{equation}
for any constant $T_{a}$. Hence, $Z^{a}(x)$ occupies similar amount
of functional volume as the residual $\Unitary(1)$ gauge symmetry
associated with Lorentz gauge fixing: $\partial_{\mu}A^{\mu}=0.$
Eq.~(\ref{eq:establishreality-1}) states that the theory of the
local field $\theta^{a}(x)$ is related to the theory of a non-local
operator if viewed from the $A_{\mu}(x)$ perspective. On the other
hand, the exact nature of the relationship depends on how the data
$\{Z^{a}(y),Y_{0}\}$ are handled in the partition function. This
will be discussed in Sec.~\ref{sec:Path-Integral}

The substitution of Eq.~(\ref{eq:thetaa}) into Eq.~(\ref{eq:matrix})
gives us explicitly the relationship between $A_{\mu}$ and $G_{\,\,\,\,\,\,\gamma}^{\alpha}$:
\begin{equation}
G_{\,\,\,\,\,\,\beta}^{\alpha}(y)=\exp\left[i\sum_{a=1}^{N}\left(\int_{Y_{0}}^{y}dz^{\mu}(H^{a})_{\,\,\,\,\,\,\mu}^{\lambda}A_{\lambda}\left(x\left(z,y\right)\right)+Z^{a}\left(y\right)\right)H^{a}\right]_{\,\,\,\,\,\,\beta}^{\alpha}.
\end{equation}
Hence, when expressed in terms of $A_{\mu}(x)$, this theory looks
manifestly like a non-local theory just as in the case of the Wilson
line field. However, when expressed in terms of $\theta^{a}(y)$ without
reference to $A_{\mu}$, the theory is manifestly local. The two seemingly
conflicting viewpoints will be reconciled later, where we will see
that symmetries of the theory in terms of $\theta^{a}(y)$ will cause
the theory to be insensitive to $Y_{0}$ and $Z^{a}$, eliminating
most of the non-locality. At the same time, it is interesting that
field operators formed out of $\theta^{a}(y)$ exist which are multilocal
at \emph{finite} number of discrete points, do not depend on $Y_{0}$
or $Z^{a}$, but represent a sum of \emph{infinite} number of $A_{\mu}$
operators (\emph{i.e.}, an integral over $A_{\mu}$):

\begin{align}
\mathcal{O}_{a}(y,T_{a}) & \equiv\int_{y}^{y+T_{a}\psi_{(a)}}dz^{\mu}(H^{a})_{\,\,\,\,\,\,\mu}^{\lambda}A_{\lambda}(x(z,y+T_{a}\psi_{(a)}))\\
 & =\theta^{a}(y+T_{a}\psi_{(a)})-\theta^{a}(y)\,\,\,\,\,\,\,\,\,\mbox{no sum over }a\label{eq:twopoints}
\end{align}
Hence, it is interesting that BTGT allows us to collapse an integral
of local fields into evaluation of local fields at two points. It
is beyond the scope of this paper to see if this feature lends itself
to an interesting description of holography (see e.g.~\cite{Aharony:1999ti,Nastase:2007kj}).

\section{\label{sec:Path-Integral}Partition Function}

Now that we have identified the field that we wish to use to describe
the gauge theory, we need to construct the partition function. What
we can do to construct the partition function is to start with the
$A_{\mu}$ theory and make a change of variables to the $\theta^{a}$
theory. After the construction, we can eliminate the starting point
of the $A_{\mu}$ and give the path integral construction rules just
in terms of $\theta^{a}$. However, we will see that we need to impose
a new symmetry to carry out this program.

The procedure to start from the $A_{\mu}$ theory is as follows: 
\begin{enumerate}
\item Start with an ordinary gauge theory functional measure and ordinary
$\xi$-gauge fixing: 
\begin{equation}
\mathcal{Z}_{1}=N_{\xi}\int Dg|\det\square|\int DAD\phi D\phi^{*}e^{i\left(S+S_{gf}\right)},\label{eq:usualgaugepathintegral}
\end{equation}
in which 
\begin{equation}
S_{gf}=\frac{-1}{2\xi}\int d^{4}x(\partial^{\mu}A_{\mu})^{2}
\end{equation}
and $S$ contains the matter field and ordinary gauge invariant combination
of $A_{\mu}$. 
\item Make a change of variables using Eq.~(\ref{eq:establishreality-1}):
\begin{equation}
\mathcal{Z}_{1}=N_{\xi}\int Dg|\det\square|\mathcal{J}\int D\theta_{nz}D\phi D\phi^{*}e^{i\left(S+S_{gf}\right)},
\end{equation}
in which 
\begin{eqnarray}
\mathcal{J} & = & \left|{\rm det}\left[\frac{\delta A_{\mu}(x)}{\delta\theta_{nz}^{a}(y)}\right]\right|\\
 & = & \left|\det\left[(H^{a})_{\,\,\,\,\,\,\mu}^{\alpha}\frac{\partial}{\partial x^{\alpha}}\delta^{(4)}(x-y)\right]\right|,
\end{eqnarray}
and $\theta_{nz}^{a}$ stands for functions which are not annihilated
by 
\begin{equation}
(H^{a})_{\,\,\,\,\,\mu}^{\alpha}\frac{\partial}{\partial x^{\alpha}}.\label{eq:partialageneral}
\end{equation}
(Note that if we do not separate the zero modes out, then we would
obtain $\mathcal{J}=0$.) However, it is difficult to restrict the
integration to $\theta_{nz}$ and it is worthwhile to find a way to
include the zero modes of Eq.~(\ref{eq:partialageneral}). One way
to do this is to multiply by $D\theta_{z}$ which integrates over
zero modes: 
\begin{eqnarray}
\mathcal{Z}_{2} & = & \int D\theta_{z}\mathcal{Z}_{1}\label{eq:increasemeasure}\\
 & = & \mathcal{N}\int D\theta D\phi D\phi^{*}e^{i\left(S[\theta,\phi,\phi^{*}]+S_{gf}[\theta]\right)}
\end{eqnarray}
\begin{equation}
\mathcal{N}\equiv N_{\xi}\int Dg|\det\square|\mathcal{J}.
\end{equation}
This should be as harmless as multiplying by the residual gauge degrees
of freedom in the Feynman gauge. This is the main difference between
the ordinary gauge theory and the BTGT theory, and it most likely
will not show up in perturbative computations, just as the residual
gauge degree of freedom in Feynman gauge does not destroy perturbation
theory. 
\end{enumerate}
Hence, we now have the partition function $\mathcal{Z}_{2}$ describing
the theory of $\theta^{a}$ and $\phi$.

At this point, we can forget that we started with the $A_{\mu}$ theory
and construct the theory of $\theta^{a}$ and $\phi$ using the following
procedure: 
\begin{enumerate}
\item Define the partition function in $\xi$-gauge as 
\begin{equation}
\boxed{\mathcal{Z}_{3}=\int D\theta D\phi D\phi^{*}\exp\left[i\left(S[\theta,\phi,\phi^{*}]-\frac{1}{2\xi}\int d^{4}x\left[\sum_{a}(H^{a})_{\,\,\,\,\,\mu}^{\alpha}\partial^{\mu}\partial_{\alpha}\theta^{a}\right]^{2}\right)\right]}
\end{equation}

\item Choose $S$ such that it is invariant under the usual Lorentz-invariant
local field theory symmetry and the following two additional symmetries:

\begin{enumerate}
\item Gauge invariant under the $\Unitary(1)$ transformations: 
\begin{equation}
\theta^{a}(x)\rightarrow\theta^{a}(x)-\Lambda(x)\label{eq:gaugeinvariant1}
\end{equation}
\begin{equation}
\phi(x)\rightarrow e^{i\Lambda(x)}\phi(x)\,\,\,\,\,\,\,\,\phi^{*}(x)\rightarrow e^{-i\Lambda(x)}\phi^{*}(x)\label{eq:gaugeinvariant2}
\end{equation}

\item Invariant under a lower dimensional functional shift transformation:
\begin{equation}
\boxed{\theta^{a}(x)\rightarrow\theta^{a}(x)+Z^{a}(x)}\label{eq:residualgauge}
\end{equation}
where 
\begin{equation}
(H^{a})_{\,\,\,\,\mu}^{\alpha}\frac{\partial}{\partial x^{\alpha}}Z^{a}(x)=0\,\,\,\,\,\,\,\mbox{no sum over }a.\label{eq:lowerdimgaugesym}
\end{equation}
This is manifestly a local symmetry without gauge fields. 
\end{enumerate}
\end{enumerate}
The gauge symmetry conditions Eqs.~(\ref{eq:gaugeinvariant1}) and
(\ref{eq:gaugeinvariant2}) in item 2 lead to the usual gauge couplings
(but in terms of $\theta^{a}$) once one is guaranteed that $\theta^{a}$
only comes in the package of $A_{\mu}(\theta^{a}(y),y)$ (\emph{i.e.},
through Eq.~(\ref{eq:establishreality-1})). As we explicitly check
in the next section, this packaging is partly enforced by the local
symmetry Eq.~(\ref{eq:residualgauge}). Furthermore, this local symmetry
is very important in that it eliminates gauge theory destabilizing
terms $\Delta\mathcal{L}_{1}$ of the form 
\begin{eqnarray}
\Delta\mathcal{L}_{1} & = & \frac{\mu^{2}}{16}|\phi|^{2}\left({\rm tr}G\right)\left({\rm tr}G^{-1}\right)\label{eq:gaugeinvariantbutnotBTGTinvariant}\\
 & \approx & \mu^{2}|\phi|^{2}\left(1-\frac{3}{16}\sum_{a}(\theta^{a})^{2}+\frac{1}{16}\sum_{b\neq c}\theta^{b}\theta^{c}+O(\theta^{4})\right),
\end{eqnarray}
which is gauge invariant in the sense of Eqs.~(\ref{eq:gaugeinvariant1})
and (\ref{eq:gaugeinvariant2}), but not Eq.~(\ref{eq:residualgauge}).
Note that this local symmetry also forbids global charge violating
terms such as 
\begin{equation}
\Delta\mathcal{L}_{2}=\frac{\mu^{2}}{16}\left[\phi^{2}\left({\rm tr}G\right)^{2}+h.c.\right],\label{eq:chargeviolating}
\end{equation}
which means that the theory inherits the global charge conservation
as an accidental symmetry just as in ordinary gauge theories once
the ordinary gauge symmetry condition is imposed. We note that as
long as the measure is chosen such that $D\theta$ is integrated over
an unrestricted function space, Eq.~(\ref{eq:residualgauge}) is
not anomalous, at least in flat space.

Before closing this section, it is important to emphasize that Eq.~(\ref{eq:residualgauge})
is a symmetry that is new and intrinsic to BTGT. This symmetry's origin
is in the derivative operator appearing in Eq.~(\ref{eq:establishreality-1}),
which does not have an analog in ordinary gauge theories. As alluded
to in Eq.~(\ref{eq:zeromodedef}), this symmetry is the main reason
why the integration origin $Y_{0}$ and the arbitrary function $Z^{a}(y)$
appearing in Eq.~(\ref{eq:thetaa}) are not meaningful. (More discussion
of this in terms of translational invariance is given in Appendix
\ref{sec:Spacetime-translation-symmetry}). This in turn means that
even though naively $G_{\,\,\,\,\,\beta}^{\alpha}(x)$ when expressed
in terms of the gauge field (\emph{i.e.} Eq.~(\ref{eq:thetaa}))
seems to be just as non-local as a Wilson line operator, it is not.
At the same time, as shown in Eq.\ (\ref{eq:twopoints}), $\theta^{a}(x)$
has a different degree of locality when compared to the gauge field
$A_{\mu}(x)$, since two points are effectively mapped to an integral
of $A_{\mu}(x)$ (\emph{i.e.}, a sum over an infinite number of points).
Incidentally, we call the shift function $Z^{a}(x)$ a lower-dimensional
function because Eq.~(\ref{eq:lowerdimgaugesym}) implies Eq.~(\ref{eq:zeromodedef}).

One naive downside of this construction is that power-counting is
more difficult because $\theta^{a}$ is a dimensionless variable.
Unlike a sigma model parameterization where the kinetic term for the
analog of $\theta^{a}$ is of the form $(\partial_{\mu}\theta)^{2}$
which would allow $\theta$ to acquire dimension upon canonical normalization,
the $\theta^{a}$ kinetic term is quartic in derivatives. However,
due to the new local symmetry Eq.~(\ref{eq:residualgauge}), $\theta^{a}$
always enters with derivatives. Hence, there does not seem to be real
harm done to bottom up model constructions by the loss of power counting.
Incidentally, we show in section \ref{sub:Hamiltonian-is-bounded}
that even though the higher derivative nature of the theory might
seem to imply that we should worry about the stability of the theory
(Ostrogradsky instability\cite{Ostrogradsky:1850fid}), the theory
is stable as the Hamiltonian is bounded from below. This stability
is related to the fact that the additional local symmetry of Eq.~(\ref{eq:residualgauge})
makes the Hamiltonian identical to ordinary gauge theories.

\section{\label{sec:Elementary-computation}Elementary computation}

Let us consider a simple example theory and compute a simple scattering
process as a basic check of the formalism. Consider a scalar field
$\phi$ charged under a $\Unitary(1)$ gauge charge $e$. The quadratic
term for the $\phi$ field that is invariant under the global $\Unitary(1)$
subgroup is 
\begin{equation}
\Delta\mathcal{L}_{k1}=|\partial\phi|^{2}-m^{2}|\phi|^{2}.
\end{equation}
(We can of course add quartic self-interactions at the marginal operator
level, but we will omit it since we will not be using it.) As noted
in Eqs.~(\ref{eq:gaugeinvariant1-1}) and (\ref{eq:gaugeinvariant2-1}),
we have to impose a separate gauge invariance given by 
\begin{equation}
e\theta^{a}(x)\rightarrow e\theta^{a}(x)-e\Lambda(x)\label{eq:gaugeinvariant1-1}
\end{equation}
\begin{equation}
\phi(x)\rightarrow e^{ie\Lambda(x)}\phi(x)\,\,\,\,\,\,\,\,\phi^{*}(x)\rightarrow e^{-ie\Lambda(x)}\phi^{*}(x).\label{eq:gaugeinvariant2-1}
\end{equation}
as well as the new local symmetry (Eq.~(\ref{eq:residualgauge}))
\begin{equation}
e\theta^{a}(x)\rightarrow e\theta^{a}(x)+eZ^{a}(x).\label{eq:newlocal}
\end{equation}

To consider the ramifications of Eq.~(\ref{eq:newlocal}) a bit more
explicitly, consider the $\theta^{a}$ dependent terms in the Lagrangian
be a Lorentz invariant function combination $\mathcal{F}(\theta,\partial_{\mu}\theta,\partial_{\mu}\partial_{\nu}\theta,...)$,
where we can truncate the ``...'' at a finite derivative order due
to power counting, and restrict the new local gauge invariance to
imply the invariance of the Lagrangian instead of the action. The
variation in the action due to Eq.~(\ref{eq:newlocal}) is 
\begin{equation}
\delta\mathcal{F}(\theta,\partial_{\mu}\theta,\partial_{\mu}\partial_{\nu}\theta,...)=Z^{a}(x)\frac{\partial\mathcal{F}}{\partial\theta^{a}}+\partial_{\mu}Z^{a}(x)\frac{\partial\mathcal{F}}{\partial\partial_{\mu}\theta^{a}}+\partial_{\mu}\partial_{\nu}Z^{a}(x)\frac{\partial\mathcal{F}}{\partial\partial_{\mu}\partial_{\nu}\theta^{a}}+...\label{eq:variationoflagrangianfromlocalsim}
\end{equation}
where the sum over $a$ is implied. Since there are an infinite number
of constraints imposed on the finite number of terms, each of these
terms must vanish independently. This implies 
\begin{equation}
\frac{\partial\mathcal{F}}{\partial\theta^{a}}=0.
\end{equation}
The condition that the next term vanishes 
\begin{equation}
\partial_{\mu}Z^{a}(x)\frac{\partial\mathcal{F}}{\partial\partial_{\mu}\theta^{a}(x)}=0
\end{equation}
can be solved by 
\begin{equation}
\frac{\partial\mathcal{F}}{\partial\partial_{\mu}\theta^{a}}=(H^{a})_{\,\,\,\,\,\delta}^{\mu}\mathcal{V}^{\delta},
\end{equation}
in which $\mathcal{V}^{\delta}$ is a $(1\,\,\,\,0)$ Lorentz tensor.
This means that every $\partial_{\mu}\theta^{a}$ dependence in $\mathcal{F}$
must come in the form with $(H^{a})_{\,\,\,\,\,\,\delta}^{\mu}$ attached
since if there were any other solutions, then $Z^{a}$ would have
to satisfy other independent constraints.

Now, suppose the next term $\partial_{\mu}\partial_{\nu}Z^{a}(x)\frac{\partial\mathcal{F}}{\partial\partial_{\mu}\partial_{\nu}\theta^{a}}$
vanishes without $\frac{\partial\mathcal{F}}{\partial\partial_{\mu}\partial_{\nu}\theta^{a}}$
being proportional to $(H^{a})_{\,\,\,\,\,\delta}^{\mu}$ or $(H^{a})_{\,\,\,\,\,\delta}^{\nu}$.
Then we must impose a new constraint on $Z^{a}$: 
\begin{equation}
\mathcal{F}_{(b)}^{\mu\nu}\partial_{\mu}\partial_{\nu}Z^{b}(x)=0\,\,\,\,\,\,\,\,\,\,\,\mbox{no sum over }b\label{eq:nootherconstraint}
\end{equation}
where $\mathcal{F}_{(b)}^{\mu\nu}$ is a tensor. Since we do not want
to contradict the fact that the only constraint on $Z^{a}$ is Eq.~(\ref{eq:lowerdimgaugesym})
and it is otherwise arbitrary, Eq.~(\ref{eq:nootherconstraint})
can be possible if $\mathcal{F}_{(b)}^{\mu\nu}$ is antisymmetric.
However, that would imply 
\begin{equation}
\frac{\partial\mathcal{F}}{\partial\partial_{\mu}\partial_{\nu}\theta^{a}}
\end{equation}
is antisymmetric in $\mu\leftrightarrow\nu$, which is impossible
for the smooth $\theta^{a}$ relevant for perturbation theory. Similar
arguments apply for higher derivatives.

Hence, we conclude we can only write $\theta^{a}$ in the combination
of Eq.~(\ref{eq:establishreality-1}) for the Lorentz-invariant local
Lagrangian satisfying the invariance of Eq.~(\ref{eq:newlocal}).
The renormalizable dimension coupling between $\theta^{a}$ and $\phi$
that obeys Eq.~(\ref{eq:newlocal}) is 
\begin{eqnarray}
\mathcal{L}_{I} & = & -ig_{1}\phi^{*}\partial^{\mu}\phi\partial_{\alpha}\theta^{a}(H^{a})_{\,\,\,\,\,\mu}^{\alpha}+h.c.\nonumber \\
 &  & +g_{2}|\phi|^{2}\partial_{\alpha}\theta^{a}(H^{a})_{\,\,\,\,\,\mu}^{\alpha}\partial_{\beta}\theta^{b}(H^{b})^{\beta\mu}.
\end{eqnarray}
In addition, the renormalizable kinetic terms would be 
\begin{eqnarray}
\mathcal{L}_{k2} & = & c_{2}\left(\partial_{\alpha}\theta^{a}(H^{a})_{\,\,\,\,\,\mu}^{\alpha}\right)\left(\partial_{\beta}\theta^{b}(H^{b})^{\beta\mu}\right)+c_{41}\partial_{\mu}\left(\partial_{\alpha}\theta^{a}(H^{a})_{\,\,\,\,\,\nu}^{\alpha}\right)\partial^{\mu}\left(\partial_{\beta}\theta^{b}(H^{b})^{\beta\nu}\right)\nonumber \\
 &  & +c_{42}\partial_{\mu}\left(\partial_{\alpha}\theta^{a}(H^{a})_{\,\,\,\,\,\nu}^{\alpha}\right)\partial^{\nu}\left(\partial_{\beta}\theta^{b}(H^{b})^{\beta\mu}\right),
\end{eqnarray}
in which the repeated indices are summed. Imposing Eqs.~(\ref{eq:gaugeinvariant1-1})
and (\ref{eq:gaugeinvariant2-1}) on $\mathcal{L}=\mathcal{L}_{k1}+\mathcal{L}_{k2}+\mathcal{L}_{I}$
results in setting $c_{2}=0$, $c_{41}=-c_{42}=2c$ (where $c$ is
a constant determined by Coulomb scattering), $g_{1}=e$, and $g_{2}=e^{2}$.
We note that after imposing the invariance of Eq.~(\ref{eq:newlocal}),
the rest of the invariances fixing these coefficients are identical
to ordinary gauge invariance.

To simplify the computations, it is useful to go to a Lorentz frame
in which $\theta^{a}H^{a}$ is diagonal: 
\begin{equation}
\sum_{a}\theta^{a}(H^{a})_{\,\,\,\,\,\nu}^{\mu}=\sum_{a}\bar{\theta}^{a}\delta_{(a)}^{\mu}\delta_{{(a)}\nu}.
\end{equation}
In this gauge, the $\langle\theta^{\beta}\theta^{\lambda}\rangle$
analog of the $\langle A_{\mu}A_{\nu}\rangle$ propagator giving the
Feynman rule $i\eta_{\mu\nu}/(4ck^{2})=-i\eta_{\mu\nu}/k^{2}$ (where
$\eta_{\mu\nu}=\mbox{diag}(1,-1,-1,-1)$) is 
\begin{equation}
\int d^{4}xe^{ik\cdot(x-y)}\langle\bar{\theta}^{\beta}(x)\bar{\theta}^{\lambda}(y)\rangle=\frac{i\frac{\eta^{\beta\lambda}}{k_{\beta}k_{\lambda}}}{4ck^{2}}\,\,\,\,\,\,\,\,\mbox{no sum}\label{eq:framefixedpropagator}
\end{equation}
where one can count the minus signs as $(i)^{2}$ coming from $k^{\beta}k^{\lambda}$
and an extra minus sign from integrating by parts one of the factors
$\partial_{\delta}\bar{\theta}^{\delta}$ to obtain the quartic differential
operator to invert. The cubic and quartic vertices are 
\begin{equation}
\frac{\partial}{\partial\phi}\frac{\partial}{\partial\phi^{*}}\frac{\partial}{\partial\bar{\theta}^{\gamma}}i\mathcal{L}_{I}|=\left[p+k\right]^{\gamma}q_{\gamma}e\,\,\,\,\,\,\,\,\,\mbox{no sum}
\end{equation}
\begin{equation}
\frac{\partial}{\partial\phi}\frac{\partial}{\partial\phi^{*}}\frac{\partial}{\partial\bar{\theta}^{\gamma}}\frac{\partial}{\partial\bar{\theta}^{\lambda}}i\mathcal{L}_{I}|=-2iq_{\lambda}r^{\lambda}\delta_{\,\,\,\,\gamma}^{\lambda}e^{2}\,\,\,\,\,\,\,\,\,\mbox{no sum}
\end{equation}
according to the usual prescription. The noncovariant notation here
comes from having made a frame choice that $H^{a}$ are diagonal matrices.
For example, a more manifestly covariant tree-level propagator is
\begin{equation}
\int d^{4}xe^{ik\cdot(x-y)}\langle\theta^{b}(x)\theta^{a}(y)\rangle=\frac{i}{4c}\frac{\delta_{\,\,\,\,\, a}^{b}}{(H^{b})_{\,\,\,\,\,\delta}^{\mu}k_{\mu}k_{\nu}(H^{a})_{\,\,\,\,\,\gamma}^{\nu}\eta^{\delta\gamma}}\frac{1}{k^{2}}\label{eq:covariant-propagator}
\end{equation}
which reverts to Eq.~(\ref{eq:framefixedpropagator}) when 
\begin{equation}
\left(H^{a}\right)_{\,\,\,\,\,\,\beta}^{\alpha}=\delta_{(a)}^{\alpha}\delta_{(a)\beta}.\,\,\,\,\mbox{no sum over }a.
\end{equation}

The t-channel tree-level Coulomb scattering gives the amplitude 
\begin{equation}
i\mathcal{M}=-ie^{2}\frac{1}{4c(k_{1}-p_{1})^{2}}\left[p_{1}+k_{1}\right]\cdot\left[p_{2}+k_{2}\right]
\end{equation}
which matches the usual scalar field theory result with $c=-1/4$
as expected.

\section{\label{sec:Peculiarities-of-the}Peculiarities of the formalism}

\subsection{Charge dependent axes}

It is interesting to note that we can rewrite the covariant derivative
as an ordinary divergence acting on a composite field consisting of
$G_{\,\,\,\,\,\beta}^{\alpha}$ and a matter field $\phi_{1}$: 
\begin{equation}
\left(\frac{\partial}{\partial x^{\mu}}+iq_{1}A_{\mu}(x)\right)\phi_{1}(x)=\frac{\partial}{\partial y_{(q_{1})}^{\alpha}}\Psi_{(q_{1})\,\,\,\,\,\,\mu}^{\alpha}(x)\label{eq:derivativeonrhs}
\end{equation}
in which 
\begin{equation}
dy_{(q_{1})}^{\alpha}=G_{(q_{1})\,\,\,\,\,\mu}^{\alpha}(x)dx^{\mu}\label{eq:specialcoordinate}
\end{equation}
and 
\begin{equation}
\Psi_{(q_{1})\,\,\,\,\,\mu}^{\alpha}(x)\equiv\phi_{1}(x)G_{(q_{1})\,\,\,\,\,\,\mu}^{\alpha}(x),
\end{equation}
where there is a mismatch between the derivative variable $y_{(q_{1})}$
on the right hand side of Eq.~(\ref{eq:derivativeonrhs}) and the
argument of $\Psi_{(q_{1})\,\,\,\,\,\,\delta}^{\lambda}(x)$. We note
that since $\Psi_{(q_{1})\,\,\,\,\,\delta}^{\lambda}$ is a covariant
tensor, the tensor components in the $y$ coordinate system is different
from that in the $x$ coordinate system. Furthermore, unlike before,
we have displayed the charge assignment of the gauge group explicitly.
Hence, if the $G$ tensor is treated as a spacetime axis, then there
are as many axes in spacetime as there are number of different charges.
On the other hand, there is only one set of basis tensor fields $\theta^{a}$
that decomposes all of the axes, at least when matching to standard
gauge theories.

\subsection{\label{sub:Hamiltonian-is-bounded}Hamiltonian is bounded from below}

It is well known that higher derivative theories generally exhibit
an instability associated with the Hamiltonian being unbounded from
below (for a review, see e.g.~\cite{Hawking:2001yt,Woodard:2006nt,Antoniadis:2006pc,Chen:2012au,Salvio:2015gsi}).
This instability is sometimes referred to as the Ostrogradsky instability.
Here, we will show that although BTGT is a higher derivative theory,
it has a Hamiltonian that is bounded from below. This can be partially
explained by the novel local symmetry Eq.~(\ref{eq:residualgauge})
which effectively eliminates the $\theta^{a}$ degree of freedom from
the action in favor of $(H^{a})_{\,\,\,\,\,\nu}^{\mu}\partial_{\mu}\theta^{a}$.

The energy density for a gauged massive scalar field is 
\begin{equation}
T_{00}=T_{00}^{(\phi)}+T_{00}^{(\theta)}
\end{equation}
where 
\begin{equation}
T_{00}^{(\phi)}=|\partial_{0}\phi+iA^{0}(\theta)\phi|^{2}+\sum_{i=1}^{3}|\partial_{i}\phi-iA^{i}(\theta)\phi|^{2}+m^{2}|\phi|^{2}
\end{equation}

\begin{equation}
T_{00}^{(\theta)}=\frac{1}{2}\sum_{i=1}^{3}\left(\partial_{0}A^{i}(\theta)+\partial_{i}A^{0}(\theta)\right)^{2}+\frac{1}{2}\sum_{l=1}^{3}\left(\sum_{m,n=1}^{3}\epsilon_{lmn}\partial_{m}A^{n}(\theta)\right)^{2}
\end{equation}
\begin{equation}
A_{\delta}(\theta)=\sum_{a}\partial_{\mu}\theta^{a}(H^{a})_{\,\,\,\,\,\,\delta}^{\mu},\label{eq:thetapackage}
\end{equation}
which is positive definite. Hence, we do not expect the Ostrogradsky
instability to arise in this theory. Again, this protection partly
comes from the novel local symmetry Eq.~(\ref{eq:residualgauge}).
As discussed around Eq.~(\ref{eq:variationoflagrangianfromlocalsim}),
other ingredients include locality and Lorentz invariance, which all
play a role in having $\theta^{a}$ come in the form of Eq.~(\ref{eq:thetapackage}).

\subsection{Computing non-local correlators}

We can in principle use the new formalism to compute non-local correlators
in novel ways. For example, consider the correlator 
\begin{equation}
\mathcal{G}_{a_{1}a_{2}}\equiv\langle\mathcal{O}_{a_{1}}(x_{1},T_{a_{1}})\mathcal{O}_{a_{2}}(x_{2},T_{a_{2}})\rangle,
\end{equation}
in which $\mathcal{O}_{a}$ are the operators defined in Eq.~(\ref{eq:twopoints}).
Note that $\mathcal{G}_{a_{1}a_{2}}$ is invariant under the local
transformations of Eq.~(\ref{eq:residualgauge}). This correlator
is easy to compute in the BTGT formalism. At tree level, it is given
by 
\begin{equation}
\mathcal{G}_{a_{1}a_{2}}=-i\int\frac{d^{4}k}{(2\pi)^{4}}\frac{e^{-ik\cdot(x_{1}-x_{2})}\delta_{a_{1}a_{2}}}{\left(k^{2}+i\epsilon\right)(H^{a_{1}})_{\,\,\,\,\,\delta}^{\mu}k_{\mu}k_{\nu}(H^{a_{2}})_{\,\,\,\,\,\gamma}^{\nu}\eta^{\delta\gamma}}\left(e^{iT_{a_{1}}k\cdot\psi_{(a_{1})}}-1\right)\left(e^{-iT_{a_{2}}k\cdot\psi_{(a_{2})}}-1\right),\label{eq:non-local-correlator}
\end{equation}
in which we have used Eq.~(\ref{eq:covariant-propagator}). This
result in the usual $A_{\mu}$ formalism corresponds to 
\begin{eqnarray}
\mathcal{G}_{a_{1}a_{2}} & = & \int_{x_{1}}^{x_{1}+T_{a_{1}}\psi_{(a_{1})}}dz_{1}^{\mu}(H^{a_{1}})_{\,\,\,\,\,\,\mu}^{\lambda}\int_{x_{2}}^{x_{2}+T_{a_{2}}\psi_{(a_{2})}}dz_{2}^{\nu}(H^{a_{2}})_{\,\,\,\,\,\,\nu}^{\beta}\times\nonumber \\
 &  & \langle A_{\lambda}(x(z_{1},x_{1}+T_{a_{1}}\psi_{(a_{1})}))A_{\beta}(x(z_{2},x_{2}+T_{a_{2}}\psi_{(a_{2})}))\rangle.\label{eq:gaugefieldcorrelator}
\end{eqnarray}
Hence, this offers a novel way to compute correlators. In the limit
$T_{a_{1}}=T_{a_{2}}=T\rightarrow0$, Eq.~(\ref{eq:non-local-correlator})
becomes 
\begin{equation}
\mathcal{G}_{a_{1}a_{2}}=-iT^{2}\int\frac{d^{4}k}{(2\pi)^{4}}\frac{e^{-ik\cdot(x_{1}-x_{2})}\eta^{a_{1}a_{2}}}{k^{2}+i\epsilon},
\end{equation}
recovering the photon propagator information. Hence, $\mathcal{G}_{a_{1}a_{2}}$
is a non-local object that in the local limit gives back the photon
propagator. It is interesting that the local limit of the fundamental
non-local Green's function%
\footnote{It is fundamental since it is invariant under the new local symmetry
of Eq.~(\ref{eq:residualgauge}) defining BTGT. %
} of BTGT is the ordinary photon Green's function.

\section{\label{sec:Conclusions}Conclusions}

In this paper, we have constructed a novel formulation for gauge theories
based on analogies with the vierbein formulation of general relativity.
For simplicity, we have focused on a simple $\Unitary(1)$ theory
in this work. This has led us to introduce a vierbein-like field $G_{\,\,\,\,\,\beta}^{\alpha}(x)$
(indicating the direction in the gauge group representation space)
that can be further decomposed (to solve constraint equations) in
terms of another set of basis tensor fields $\theta^{a}(x)(H^{a})_{\,\,\,\,\,\nu}^{\mu}$.
Unlike the Wilson line, $\theta^{a}(x)$ is a local field. The basis
tensor field $\theta^{a}(x)$ has new local symmetries given by Eq.~(\ref{eq:residualgauge})
that are important for preserving translational invariance as discussed
in Sec.~\ref{sec:Spacetime-translation-symmetry} and maintaining
stability as discussed in Sec.~\ref{sub:Hamiltonian-is-bounded}.
Intuitively, the field theory of $\theta^{a}$ contains the gauge
theory information by way of Eq.~(\ref{eq:gaugefieldcorrelator}).

There are many future research directions that are suggested by this
work. Perhaps most obviously, BTGT should be generalizable to non-Abelian
theories.%
\footnote{There are also certain technical details of the construction in this
paper that can be improved. For example, although the argument surrounding
Eq.~(\ref{eq:variationoflagrangianfromlocalsim}) is sufficient for
constructing an action only in terms of $A_{\mu}(\theta)$, it does
not address the possibility of the action having variations of a total
derivative term.%
} It would also be interesting to find practical applications for this
theory in computing non-local correlators similar to Eq.~(\ref{eq:non-local-correlator}).
The novelty in part is related to the different degree of locality
due to the higher derivative nature of this theory as noted around
Eq.~(\ref{eq:twopoints}). Loop corrections, BRST invariance, Ward
identities associated with the new local symmetry of Eq.~(\ref{eq:residualgauge})
may be interesting to explore. Instantons, sphalerons, and other non-perturbative
excitations in BTGT may be a bit different in ordinary gauge theories
since the gauge theory has been non-perturbatively modified through
the measure (see Eq.~\ref{eq:increasemeasure})). This formalism
should also be tested by embedding it into curved spacetime.

It is interesting that matter fields and gauge fields in this formalism
can be packaged in the same category of mutually dual objects in group
representation space. However, one satisfies a constraint equation
and the other does not. If there can be a way to spontaneously generate
this asymmetry starting from a even more symmetric framework, that
would open up new avenues for constructing physics beyond the SM. 
\begin{acknowledgments}
This work was supported by the DOE through grant DE-FG02-95ER40896.
DJHC thanks Lisa Everett and Aki Hashimoto for discussions related
to the general topics in this work. 
\end{acknowledgments}
\appendix

\section{\label{sec:Lower-rank-tensor}Lower rank tensor}

Instead of a rank 2 tensor as in Eq.~(\ref{eq:rank2}), suppose we
postulated a Lorentz scalar transforming under $\Unitary(1)$ as a
matter dual field representing the matter direction in representation
space. There are then not enough local functional degrees of freedom
to replace an $N$-vector field.%
\footnote{We note that the approach of \cite{Mandelstam:1962mi} effectively
has a non-local function that is a scalar: \emph{i.e.} $h_{P}(x,x_{i})=\int_{x_{i},P}^{x}dX^{\mu}A_{\mu}(X)$
where $P$ is a path. The manifest nature of the non-locality can
be seen by the fact that it is a path dependent functional and the
field strength is derived from $h_{P}$ through $h_{P+\delta P}(x,x_{i})-h_{P}(x,x_{i})=F_{\mu\nu}\sigma^{\mu\nu}$,
where $\sigma^{\mu\nu}$ represents the area of the path difference
$\delta P$. %
}

The next smallest rank to consider is 1. Suppose we choose 
\begin{equation}
G_{\gamma}(x_{1})=S_{\gamma}e^{-i\theta(x_{1})}.
\end{equation}
Because of Eq.~(\ref{eq:defofgaugepotential}), we want to solve
for $\partial_{\mu}\theta(x_{1})$ in terms of $G$ evaluated at $x_{1}$.
To this end, we can take derivatives of the general gauge transformed
object: 
\begin{equation}
e^{i\theta}(G^{-1})^{\gamma}\partial_{\mu}(G_{\gamma}e^{-i\theta})=(G^{-1})^{\gamma}\partial_{\mu}G_{\gamma}-i\partial_{\mu}\theta
\end{equation}
in which $(G^{-1})^{\gamma}G_{\gamma}\equiv1$ defines the inverse.
Evaluating this general expression at $x_{1}$ in the special gauge
frame yields 
\begin{equation}
(G^{-1})^{\gamma}\partial_{\mu}G_{\gamma}|_{x_{1}}=(\tilde{G}^{-1})^{\gamma}\partial_{\mu}\tilde{G}_{\gamma}(x_{1})-i\partial_{\mu}\theta(x_{1}).
\end{equation}
Because of Eq.~(\ref{eq:defofgaugepotential}), we conclude 
\begin{equation}
A_{\mu}(x_{1})=-i\left[(G^{-1})^{\gamma}(\partial_{\mu}G_{\gamma})|_{x_{1}}-(\tilde{G}^{-1})^{\gamma}(\partial_{\mu}\tilde{G}_{\gamma})|_{x_{1}}\right],\label{eq:vecpotatpoint-1}
\end{equation}
in which 
\begin{equation}
G_{\alpha}(x)\equiv\tilde{G}_{\alpha}(x)e^{-i\theta(x)}
\end{equation}
is the general gauge field.

We can now simplify Eq.~(\ref{eq:vecpotatpoint-1}) further by noting
that Eq.~(\ref{eq:vecpotatpoint-1}) has an additional $\tilde{\Unitary}(1)$
symmetry transformation 
\begin{equation}
G_{\gamma}\rightarrow G_{\gamma}e^{-i\theta_{2}(x)}
\end{equation}
\begin{equation}
\tilde{G}_{\gamma}\rightarrow\tilde{G}_{\gamma}e^{-i\theta_{2}(x)}
\end{equation}
that leaves Eq.~(\ref{eq:vecpotatpoint-1}) invariant. This means
we can use it to choose $\partial_{\mu}\tilde{G}_{\gamma}=0$ as follows.
First, we execute a $\tilde{\Unitary}(1)$ transform 
\begin{equation}
(G^{-1})^{\gamma}(\partial_{\mu}G_{\gamma})=(\bar{G}^{-1})^{\gamma}(\partial_{\mu}\bar{G}_{\gamma})-i\partial_{\mu}\theta_{2}\,\,\,\,\,\mbox{no sum}
\end{equation}
\begin{equation}
(\tilde{G}^{-1})^{\mu\gamma}(\partial_{\mu}\tilde{G}_{\gamma\delta})=(\bar{\tilde{G}}^{-1})^{\mu\gamma}(\partial_{\mu}\bar{\tilde{G}}_{\gamma\delta})-i\partial_{\mu}\theta_{2}\,\,\,\,\,\mbox{no sum}
\end{equation}
parametrized by a yet-to-be-determined $\theta_{2}$. We then impose
the condition 
\begin{equation}
(\bar{\tilde{G}}^{-1})^{\gamma}(\partial_{\mu}\bar{\tilde{G}}_{\gamma})=0\label{eq:gaugecondition-1}
\end{equation}
to solve for $\theta_{2}.$ This implies 
\begin{equation}
(\bar{G}^{-1})^{\gamma}(\partial_{\mu}\bar{G}_{\gamma})=(G^{-1})^{\gamma}(\partial_{\mu}G_{\gamma})-(\tilde{G}^{-1})^{\gamma}(\partial_{\mu}\tilde{G}_{\gamma}).
\end{equation}
In this $\tilde{\Unitary}(1)$ gauge fixed system, we have 
\begin{equation}
A_{\mu}(x)=-i(\bar{G}^{-1})^{\gamma}(\partial_{\mu}\bar{G}_{\gamma})
\end{equation}
where the bar indicates that we have fixed the $\tilde{\Unitary}(1)$
gauge through Eq.~(\ref{eq:gaugecondition-1}). For notational convenience,
we can simply drop the bar: \emph{i.e.} 
\begin{eqnarray}
A_{\mu} & = & -i(G^{-1})^{\gamma}\partial_{\mu}G_{\gamma}\label{eq:mapaandG-1}\\
 & = & -i\frac{G^{\gamma}}{G^{\gamma}G_{\gamma}}\partial_{\mu}G_{\gamma}\\
 & = & \frac{-i}{2}\partial_{\mu}\ln G^{\gamma}G_{\gamma}
\end{eqnarray}
which is a pure gauge configuration.

Hence, we must go to higher rank tensors for a basis tensor. The next
rank tensor is rank 2, and this is what we present in this work.

\section{\label{sec:Spacetime-translation-symmetry}Spacetime translation
symmetry }

Here we discuss one way to motivate the requirement of local symmetry
as given in Eq.~(\ref{eq:residualgauge}). Suppose we start with
a theory $S[A]$ of local $A_{\mu}(x)$ and in view of making a change
of variables to $\theta^{a}$ starting from $S[A],$ suppose we add
a non-local interaction $\Delta S$ involving $\theta^{a}(A_{\mu})$
in the form of Eq.~(\ref{eq:thetaa}) 
\begin{equation}
\Delta S=\Delta S(\theta^{a}(A_{\mu}))
\end{equation}
which is ordinary $\Unitary(1)$ gauge invariant (e.g. see Eq.~(\ref{eq:gaugeinvariantbutnotBTGTinvariant}))
but not invariant under Eq.~(\ref{eq:residualgauge}). This means
that the partition function 
\begin{equation}
\mathcal{Z}_{0}=\int DA_{\mu}D\phi D\phi^{*}e^{i\left(S+\Delta S\right)}
\end{equation}
is sensitive to $Y_{0}$ in Eq.~(\ref{eq:thetaa}). However, this
breaks spacetime translational invariance, since the interactions
have a preferred point. Hence, one way to eliminate $\Delta S$ from
the theory is to impose the local symmetry Eq.~(\ref{eq:residualgauge}).

One cannot for example try to use $\mathcal{O}_{a}(y,T_{a})$ defined
in Eq.~(\ref{eq:twopoints}) as a substitute for the $\theta^{a}(x)$
field in making a change of variables from $A_{\mu}(x)$ to obtain
a local field theory because $\mathcal{O}_{a}(y,T_{a})$ is manifestly
non-local (although translationally invariant in $y$). The local
symmetry Eq.~(\ref{eq:residualgauge}) also has the advantage of
helping to protect against the Ostrogradsky instability.

\section{\label{sec:vierbein-analogy}Vierbein analogy }

In this appendix, we explicitly list the analogy between $G_{\,\,\,\,\,\beta}^{\alpha}(x)$
formalism and the general relativistic vierbein $(e_{a})_{\mu}$ formalism,
where the index $a$ is the fictitious Minkowski space index and $\mu$
is the spacetime coordinate index. As a start, the vierbein-like field
correspondence is
\begin{equation}
G_{\,\,\,\,\,\,\beta}^{\alpha}\leftrightarrow(e_{a})_{\mu},
\end{equation}
where effectively the real and imaginary elements of $G_{\,\,\,\,\,\,\beta}^{\alpha}$
(\emph{i.e.,} the real and imaginary elements of U$(1)$ maps to SO$(2)$)
are analogs of the label $\mu$, and $(\alpha,\beta)$ labels are
the analogs of $a$. The analogy of the constraint equation is
\begin{equation}
A_{\lambda}=-i(G^{-1})_{\,\,\,\,\beta}^{\alpha}\partial_{\alpha}G_{\,\,\,\,\lambda}^{\beta}\leftrightarrow\Gamma_{\lambda\beta}^{\gamma}=(e^{a})^{\gamma}\partial_{(\lambda}(e_{a})_{\beta)}+g^{\epsilon\gamma}(e^{c})_{(\beta}\partial_{\lambda)}(e_{c})_{\epsilon}-g^{\epsilon\gamma}(e^{c})_{(\beta}\partial_{|\epsilon|}(e_{c})_{\lambda)}\label{eq:gammaande-1}
\end{equation}
\begin{equation}
g_{\alpha\beta}\equiv(e_{a})_{\alpha}(e_{b})_{\beta}\eta^{ab}.
\end{equation}
The reason why $G_{\,\,\,\,\,\beta}^{\alpha}(x)$ cannot be considered
to be analogous to an ordinary dual basis element such as a coordinate
basis object $\partial_{\mu}$ is because such objects do not carry
metric information by themselves.

\bibliographystyle{JHEP}
\bibliography{btgtpaper}

\end{document}